# Nanosecond-timescale low error switching of in-plane magnetic tunnel junctions through dynamic Oersted-field assisted spin-Hall effect

S. V. Aradhya[1,∥,✉], G E. Rowlands[1,∥,†], J. Oh[1], D. C. Ralph[1,2], R. A. Buhrman[1, ✉]

1. Cornell University, Ithaca, New York 14853, USA
2. Kavli Institute at Cornell, Ithaca, New York 14853, USA

**Abstract:**

We investigate fast-pulse switching of in-plane-magnetized magnetic tunnel junctions (MTJs) within 3-terminal devices in which spin-transfer torque is applied to the MTJ by the giant spin Hall effect. We measure reliable switching, with write error rates down to $10^{-5}$, using current pulses as short as just 2 ns in duration. This represents the fastest reliable switching reported to date for any spin-torque-driven magnetic memory geometry, and corresponds to a characteristic time scale that is significantly shorter than predicted possible within a macrospin model for in-plane MTJs subject to thermal fluctuations at room temperature. Using micromagnetic simulations, we show that in the 3-terminal spin-Hall devices the Oersted magnetic field generated by the pulse current strongly modifies the magnetic dynamics excited by the spin-Hall torque, enabling this unanticipated performance improvement. Our results suggest that in-plane MTJs controlled by Oersted-field-assisted spin-Hall torque are a promising candidate for both cache memory applications requiring high speed and for cryogenic memories requiring low write energies.

Magnetic random access memory (MRAM) controlled using spin transfer torque (STT)[1–3], using either in-plane[4–6] or perpendicularly magnetized[7,8] magnetic tunnel junctions (MTJs), holds promise for replacing existing best-in-class memory technologies in several application domains because it offers the potential for non-volatility, unlimited read and write endurance, low write energy, and low standby power. For the wide application of STT-MRAM technology it is also crucial to achieve high-speed switching with low write error rates (WERs). Currently, the fastest reliable (< $10^{-5}$ WER) switching times reported for conventional 2-terminal STT-MRAM are 35 ns for in-plane MTJs,[6] and 4 ns for perpendicular MTJs.[8] Theoretical and experimental analyses have led to skepticism about the possibility for significant improvements in switching speed and reliability, particularly for in-plane magnetized devices.[4,9] Consequently, the search for speed has led to more ambitious proposals including orthogonal spin-transfer (OST) MRAM,[10] where sub-ns switching has been demonstrated, but via a precessional non-deterministic mechanism that thus far has not allowed for competitive WERs. Here we investigate the speed and reliability of spin-orbit torque[11–15] switching in three-terminal devices[16–20] that utilize the spin Hall effect (SHE)[21,22] to achieve efficient switching of an in-plane magnetized MTJ. We demonstrate reliable (≤$10^{-5}$ WER) switching using current pulses only 2 ns long. This is faster than the best reported value for reliable switching of any previous spin-torque MRAM device[4–9] – in-plane or perpendicular – with a characteristic time scale even faster than the theoretical limit expected for in-plane-magnetized MTJs in the macrospin approximation.[9,23,24]

Figure 1a shows a schematic representation of our 3-terminal device geometry (SEM micrograph in Fig. 1b). We use a 5-nm-thick Pt spin-Hall channel to generate a spin current that impinges on a $Fe_{60}Co_{20}B_{20}$ nanomagnet free layer that is part of a magnetic tunnel junction. The reference layer of the MTJ is a FeCoB/Ru/FeCoB synthetic antiferromagnetic (the full wafer stack and fabrication technique are described in the Supporting Information). A 0.7 nm Hf spacer between the Pt and FeCoB free layer is used in order to reduce the magnetic damping, following Nguyen *et al*[18]. We report data from three devices with different aspect ratios for the MTJ: a low aspect ratio ('LA', dimension 190 × 110 nm$^2$, aspect ratio 1:1.7, coercivity, $H_c$, 14 Oe), medium ('MA', 190 × 75 nm$^2$, 1:2.5, 30 Oe) and high ('HA', 190 × 45 nm$^2$, 1:4.2, 54 Oe). In the free layer of these MTJs, the aspect ratio determines the in-plane anisotropy and therefore the thermal stability factor, $\Delta$, which is the ratio of the energy barrier ($E_b$) for switching normalized by the thermal energy ($k_BT$). The MTJs are patterned by electron-beam lithography as rounded rectangular features on top of a 335 nm wide Pt/Hf channel, with a channel resistance for all devices of 1.05 k$\Omega$. The resistance-area product of the MTJ barrier is ~190 $\Omega$-$\mu m^2$ (see Supporting Information for measurement details). All the measurements we report were performed at room temperature.

Figure 1c shows magnetic-field-driven hysteresis curves for the devices with different aspect ratios. There is a residual dipole field ($H_{off}$ = 25, 62, 65 Oe for LA, MA and HA devices) due to slight imperfection in balancing the synthetic antiferromagnetic layer, that causes the centers of the hysteresis curves to be shifted from zero; the data in Fig. 1c are plotted relative to this offset. The parallel-state (P) MTJ resistances are 13.1 k$\Omega$, 14.3 k$\Omega$, and 21.7 k$\Omega$, whereas the anti-parallel (AP) state resistances are 19.9 k$\Omega$, 29.3 k$\Omega$ and 45.6 k$\Omega$ for the LA, MA and HA devices, respectively. Consequently the tunneling magnetoresistance (TMR) is ~110% for MA and HA devices, with a lower value (52%) for the LA device. The P-state resistance of the LA device is higher than expected based on the resistance-area product of the MA and HA devices (Fig. 1d, also see Fig. S1 in Supporting Information). We therefore ascribe the reduced TMR of the LA device to a greater degree of spatial non-uniformity in its magnetic state, so that the P state resistance in the LA device is not fully saturated, due to a weaker shape anisotropy.

In order to obtain quantitative measurements of the spin Hall effect in these devices, we first conduct *dc* switching experiments (Fig. 2a). These are performed using an external offset field bias ($H_{ext}=-H_{off}$) to center the hysteresis loops (as in Fig 1c). The dependence of the critical switching current density on the ramping rate of the current density ($\dot{J}$) allows us to obtain the critical switching current density[25] in the absence of thermal fluctuations ($J_{c0}$) and the thermal stability factor ($\Delta = E_b/k_B T$, at room temperature):

$$\langle J_c \rangle = J_{c0} \left(1 + \frac{1}{\Delta} \ln\left[\left|\frac{j}{J_{c0}}\right| \tau_0 \Delta\right]\right) \qquad \text{Eq. 1}$$

where $\tau_0$ is the thermal fluctuation time, taken to be 1 ns[25]. Table 1 summarizes the *dc* switching characteristics measured for the three devices.

To explore the device performance in the fast switching regime, we perform measurements of switching probability as a function of pulse voltage and pulse duration (see Supporting Information for illustrative examples of the results of such measurements). Interpolating from these measurements, we can extract the pulse durations that result in 50% switching probability for each pulse voltage used. Figure 2 shows the data and fit to these values using the macrospin model relation[23]:

$$V(\tau) = V_{c0}\left(1 + \frac{t_{c0}}{\tau}\right) \qquad \text{Eq. 2}$$

Here $V_{c0}$ is the critical pulse switching voltage and $t_{c0}$ is the critical pulse switching duration, both defined at the 50% switching probability point. $t_{c0}$ represents the timescale needed for a sample biased at $V_{c0}$ to absorb the amount of angular momentum required for fast switching, assuming no loss to damping.[24]

The voltage scales for switching, $V_{c0}$, are the same for the three aspect ratios within experimental error (Table 1). This validates our understanding that the energetics of the switching are determined by the strength of the spin Hall effect and the geometry of the Pt channel in the pulse switching regime, both of which are the same for all three devices in this report. On the other hand, the dynamics of the switching shows surprising results, which vary in detail with the aspect ratio. First, in all cases the results indicate a remarkably fast timescale, $t_{c0} \leq \sim 1 \text{ ns}$, considerably less than the $\gg 1$ ns timescale expected[23,24] from the anti-damping switching mechanism for an in-plane magnetized free layer in the rigid domain approximation.[9] In addition, the values obtained for $t_{c0}$ show a clear dependence on the device aspect ratio with the low aspect ratio device having the fastest observed switching speed. Finally, there appears to be a growing asymmetry between the P-to-AP and AP-to-P polarity switching speeds as the aspect ratio is reduced.

To understand the origin of this unexpected but technologically important speed-up in the switching speeds, we perform zero-temperature micromagnetic simulations of a representative device to capture the behavior during switching (see movies of simulations in Supporting Information). Our simulations indicate that the Oersted field generated by the current flowing in the Pt/Hf channel plays a key role in assisting the switching process that is driven by the spin Hall torque. It is important to note that the Oersted field in our 3-terminal spin Hall devices is oriented differently than in conventional 2-terminal STT-MRAM, where the field is circularly symmetric about the device since the current flows through the MTJ itself. In the 3-terminal geometry, the Oersted field is approximately uniform and in-plane. The strength of the Oersted field (~1 kA/m at a current density of $4 \times 10^{11}$ A/m² in the Pt/Hf channel)[26] can become comparable, and is opposite in direction, to the anisotropy field of the free layer (given the sign of the spin Hall effect in Pt). Still, it is surprising that the Oersted field should have a large effect on the switching within a macrospin picture, because for anti-damping spin-torque switching the strength of the anisotropy field should have minimal impact on the critical current. The switching trajectory of the magnetization, at least within the context of a macrospin, rigid-domain approximation, should also remain largely unaffected by the magnitude of the anisotropy field. Nevertheless, in the micromagnetic simulations we observe a striking difference in the switching mechanism depending on whether the Oersted field is turned on or off. In Fig. 3 the a→b→d sequence shows micromagnetic snapshots of the AP-to-P switching process for a simulation with no Oersted field; highly non-uniform micromagnetic states are generated during the switching process. This is qualitatively similar to the excitation of short-wavelength spin wave modes observed in simulations of some conventionsl STT devices.[27] However this non-uniformity is in marked contrast to the AP-to-P simulation with the Oersted field (a→c→d in Fig. 3), where the switching process progresses through significantly more uniform states. A similar distinction is also observed in the P-to-AP switching without (d→f→a) and with (d→e→a) the Oersted field. Quantitatively, the simulations indicate that the switching is completed faster with the Oersted field, especially in the P-to-AP polarity in this particular simulation. In addition, the switching is seen to start immediately upon the application of current (t=0 in the simulations), which suggests the lack of any extended buildup of precessional amplitude, or incubation time, especially since the simulations are performed at 0 K temperature. We note that the incubation time has remained a major technological limitation factor for high-speed switching of in-plane STT-MRAM devices.[10,28] Based on these observations from the micromagnetic simulations, we conclude that the fast switching is enabled by the combination of three factors: 1) The micromagnetic curvature of the free-layer magnetization that ensures a non-zero initial torque; 2) the suppression of higher order spin-wave modes in the magnetization by the Oersted field that would otherwise hinder the completion of the reversal; and 3) the avoidance of macrospin-type stagnation points due to the non-uniformities in the micromagnetic states during the switching process.

While the limited-statistics pulse voltage and duration sweeps such as shown in Fig. 2 and the Supporting Information are routinely used to report the existence of high-speed switching,[10,15,20,28] a much more rigorous test of switching reliability is required to demonstrate feasibility for technological applications. We have tested the reliability of our 3-terminal spin Hall devices by measuring WER statistics during up to $10^5$ switching attempts for each pulse duration and pulse voltage of interest. Figure 4a,b show the measured WERs with 5 ns and 2 ns pulse durations, respectively, for the three devices. Three key results are immediately apparent from these plots. First, the WERs for 5 ns pulse durations demonstrate single-exponential scaling down to WERs of $10^{-5}$ for all three devices, indicating that the micromagnetic switching trajectories are highly reliable and scale very favorably with the applied pulse voltage. Second, the WER scaling trend highlights a significant interplay between the Oersted field and the anisotropy field scale; while all three devices exhibit fast reliable switching, the lower the coercive field the greater the effect of the Oersted field in reliably speeding the reversal. Finally, the data demonstrate that the WER with a 2 ns pulse can be driven below $10^{-5}$, most clearly in the MA device. We do observe multi-exponential features at low WERs, especially for the HA and LA devices, for either P-to-AP and AP-to-P polarities, reminiscent of the 'low probability bifurcated switching' and back-hopping mechanisms discussed by Min *et al*.[4] However, we emphasize that the WER data presented here are for a pulse duration of 2 ns, which is an order of magnitude shorter than the 50 ns pulse durations that were explored by Min *et al*. Quantitatively, this 2 ns timescale precludes many of the explanations for the multi-exponential behavior based on macrospin-type switching mechanisms. We therefore conclude that this behavior in our devices stems instead from the rich micromagnetic switching mechanism at these previously unexplored speeds. Specifically, the multi-exponential features are likely due to a particular device's atomic-scale edge roughness, pinning, and any local non-uniformities of the free layer, and can therefore can be further optimized for improved performance in the 2 ns regime.

We will refer to this previously unexplored and technologically attractive reversal mechanism as "dynamic Oersted field assisted spin-Hall effect" (DOFA-SHE) switching. Based on the insight from the micromagnetic simulations, we conclude that the fast and reliable switching we measure is a general consequence of the in-plane Oersted field orientation present in the three-terminal geometry. As discussed above, although there is clearly a correlation of the degree of enhancement of the reversal speed with the coercivity of the device, the effect cannot simply be attributed to the Oersted field overcoming the coercive field. This conclusion gains further support from the fact that we have also performed fast pulse switching experiments with a HA MTJ on a Ta spin-Hall channel that has a larger, but negative spin-Hall angle ($\approx$ -0.15 *vs.* +0.08 for our Pt devices), This change in sign results in the Oersted field pointing along the anisotropy field during switching. Despite this change in direction we find that the Ta device also has sub-ns $t_{c0}$ for both AP-to-P and P-to-AP switching, with a similar scaling of the WER as the Pt HA device (see Supporting Information). We emphasize that the Oersted field does not have any detrimental effect on the long timescale thermal stability of the devices because the field is only present during the pulse that drives switching. We anticipate that the Oersted field can be engineered to optimally assist the switching of nanomagnets of a desired thermal stability by optimizing the spin-Hall channel's geometry, resistivity, spin-Hall torque efficiency and spin diffusion length.

In summary, we have established DOFA-SHE-switched in-plane-magnetized three-terminal MTJs as an attractive architecture to achieve highly reliable magnetic switching for pulse times down to 2 ns or potentially shorter. This mechanism does not require an external magnetic field to make the switching deterministic, one of the difficulties facing the development of perpendicularly-magnetized MTJs switched using spin-orbit torques. The 3-terminal DOFA-SHE geometry also has additional advantages over conventional 2-terminal STT-MRAM in that the read and write current pathways are separate, so that the 3-terminal devices allow for arbitrarily high TMRs to minimize read times as well as to reduce read disturbs (since large currents do not flow through the MTJ itself). The beneficial speed-up of switching due to the Oersted field in the three-terminal geometry not only allows the SHE switching of in-plane MTJs to be faster than demonstrated for any other magnetic memory geometry, but it also opens up new avenues for optimizing device performance in terms of data retention versus write speed. In particular, our results suggest that the non-volatile nature of magnetic memories can now be fully harnessed for both long term data retention applications (requiring large $\Delta$), as well as for fast switching applications (requiring small $t_{c0}$) where data retention is not a primary concern. Finally, DOFA-SHE might prove attractive for cryogenic memory applications where the thermal stability of small (~1) aspect ratio MTJs is increased due to the low temperatures, thereby enhancing the relative role of the Oersted field from the spin-Hall channel.

**Associated Content:**

**Supporting Information:** Fabrication, measurement and simulation details; magnetic property characterization; pulse switching probability measurements; movies of simulated micromagnetic pulse switching trajectories. This material is available free of charge via the Internet at http://pubs.acs.org.

**Author Information:**

**Corresponding authors:** E-mail: (S.V.A) sva24@cornell.edu and (R.A.B) buhrman@cornell.edu

**Author Contributions:** S.V.A. fabricated the devices, performed magnetic characterizations and designed the experiments with the guidance of R.A.B. and D.C.R. G.E.R. performed the micromagnetic simulations and constructed the pulse switching instrumentation. J.O. performed the electrical measurements, with the help of S.V.A and G.E.R. S.V.A. wrote the manuscript with feedback from all authors. ∥S.V.A and G.E.R. contributed equally to this work. †Present address for G.E.R.: Raytheon BBN Technologies, Cambridge, Massachusetts 02138, USA.

**Notes:** The authors declare no competing financial interest.

**Acknowledgements:**

The authors thank Canon ANELVA for materials deposition, C. Jermain for help with FMR characterization, M.-H. Nguyen for help with the film stack development, and P. G. Gowtham and G. D. Fuchs for discussions. This work was supported in part by the Department of Defense (DoD) Agency-Intelligence Advanced Research Projects Activity (IARPA) through the U.S. Army Research Office under Contract No. W911NF-14-C-0089. The content of the information does not necessarily reflect the position or the policy of the Government, and no official endorsement should be inferred. Additionally, this work was supported by the NSF/MRSEC program (DMR-1120296) through the Cornell Center for Materials Research, and by the NSF (Grant No. ECCS-1542081) through use of the Cornell Nanofabrication Facility/National Nanofabrication Infrastructure Network.

**Table 1.** Critical *dc* and pulsed switching parameters of the three-terminal devices. The dc switching parameters are averaged between AP-to-P and P-to-AP polarities, as these quantities are within experimental error. The pulse switching experiments reveal non-trivial asymmetries in AP-to-P and P-to-AP switching dynamics.

|  | Device | LA | MA | HA |
|---|---|---|---|---|
| **dc** | $J_{c0}$ [×10$^{11}$ A/m$^2$] | 3.1 ± 0.3 | 4.4 ± 0.3 | 4.0 ± 0.3 |
|  | $\Delta$ | 36 ± 2 | 44 ± 3 | 54 ± 5 |
|  | $H_c$ [kA/m] | 1.11 | 2.44 | 4.30 |
| **pulse** | $V_{c0}$ (AP-P) [V] | 0.58 ± 0.05 | 0.62 ± 0.05 | 0.61 ± 0.05 |
|  | $t_{c0}$ (AP-P) [ns] | 0.43 ± 0.07 | 0.65 ± 0.09 | 1.00 ± 0.15 |
|  | $V_{c0}$ (P-AP) [V] | 0.59 ± 0.02 | 0.61 ± 0.04 | 0.59 ± 0.02 |
|  | $t_{c0}$ (P-AP) [ns] | 0.18 ± 0.02 | 0.56 ± 0.06 | 1.18 ± 0.07 |

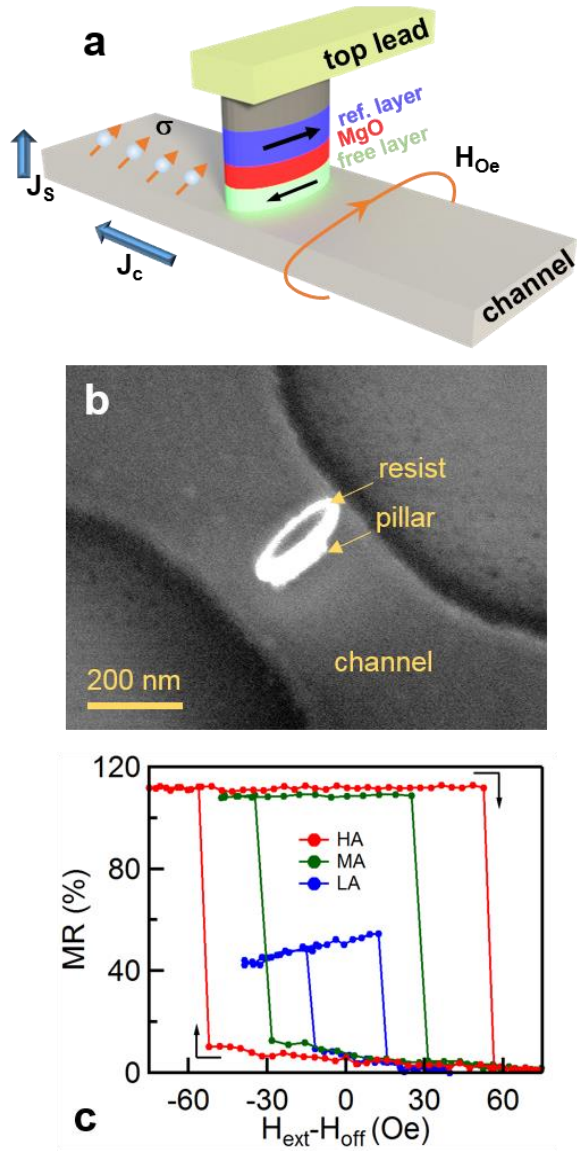

**Figure 1.** (a) Schematic of the device with directions of charge current $J_e$ and spin current $J_s$ as well as the spin accumulation **σ**. The Oersted field wraps around the Pt channel and opposes the anisotropy field in the nanomagnet during switching. (b) SEM micrograph of a representative MTJ and the spin-Hall channel obtained before top leads deposition. (c) Easy axis hysteresis loops show differences in coercive field corresponding to the aspect ratio of the MTJs.

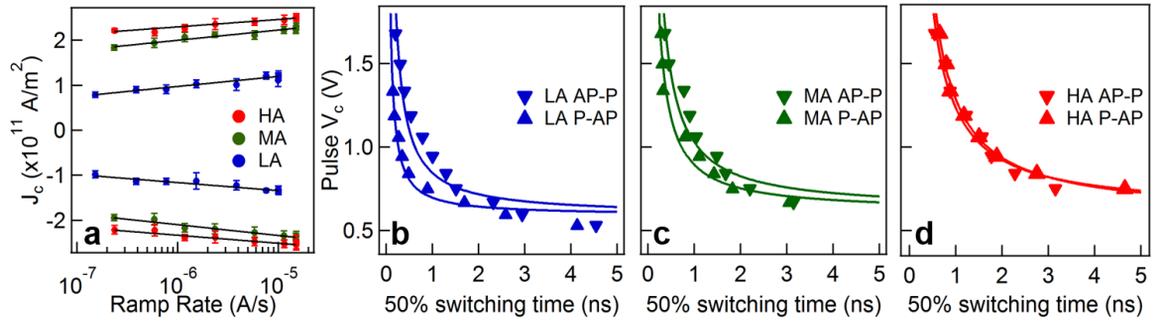

**Figure 2.** (a) dc switching current densities measured with a range of current ramp-rates for the three devices. (b,c,d) Pulsed voltage amplitudes required to achieve 50% probability of switching for a given pulse length. Lines in are are fits to Eq. 1, and in b,c,d are fits to Eq. (2).

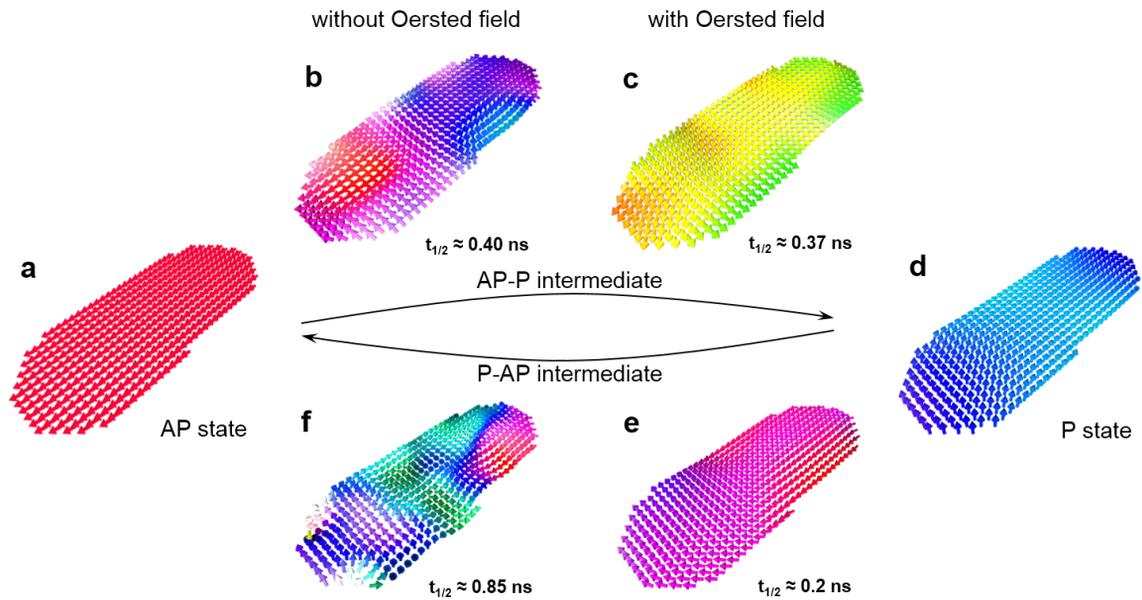

**Figure 3.** In the absence of an Oersted field, the switching mechanism for both AP-to-P (a➔b➔d) and P-to-AP (d➔f➔a) are dominated by a highly non-uniform micromagnetic intermediate states. In contrast, switching in the presence of the Oersted field proceeds through near-uniform intermediate states. The time required to complete the switching process is also significantly shorter in the presence of the Oersted field, for both AP-to-P (a➔c➔d) and P-to-AP (d➔e➔a) polarities. The intermediate states are representative snapshots taken near the halfway time ($t_{1/2}$) of the respective switching simulation.

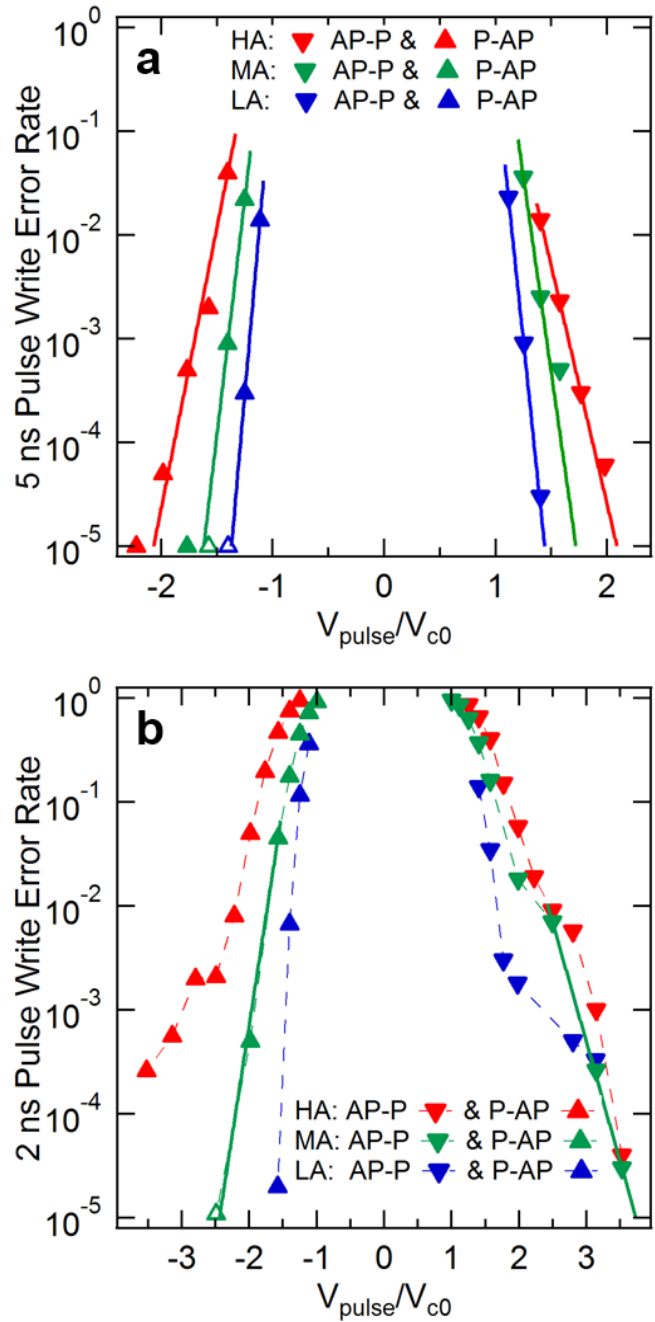

**Figure 4.** WERs for (a) 5 ns pulses and (b) 2 ns pulses. Both panels show WERs for the three devices with different aspect ratios as a function of normalized pulsed voltage. Open triangles represent data points where the measured error was zero. Dashed lines connecting the data points are provided as a visual guide. Solid lines are single-exponential fits that allow for estimation of the voltages needed to achieve error rates of $10^{-5}$.

**Table of Contents Graphic:**

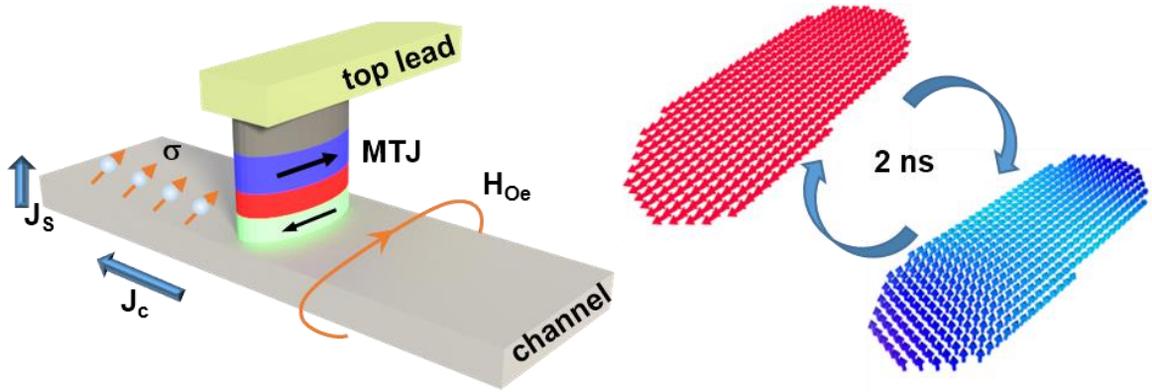

# Nanosecond-timescale low error switching of in-plane magnetic tunnel junctions through dynamic Oersted-field assisted spin-Hall effect

S. V. Aradhya[1], G E. Rowlands[1], J. Oh[1], D. C. Ralph[1,2], R. A. Buhrman[1]

1. Cornell University, Ithaca, New York 14853, USA

2. Kavli Institute at Cornell, Ithaca, New York 14853, USA

## Supporting Information

## Table of Contents

## 1. Materials and fabrication.

The magnetic multilayers are deposited onto thermally oxidized high-resistivity ($\varrho > 10{,}000$ $\Omega$cm) Si wafers using DC and RF magnetron sputtering by Canon ANELVA, Inc. Our multilayers consist of (thicknesses in nm, FeCoB≡Fe$_{60}$Co$_{20}$B$_{20}$) || SiOx | Ta(1) | Pt(5) | Hf(0.7) | FeCoB(1.6) | MgO | FeCoB(1.2) | Ta(0.2) | FeCoB(1.2) | FeCo(1) | Ru(0.85) | FeCo(2.5) | IrMn(7) | Ru(4), where the FeCoB(1.6) constitutes the magnetic free layer upon which the spin-Hall spin torque acts. The rest of the FeCoB and FeCo layers act as a synthetic antiferromagnet and the IrMn layer provides pinning through exchange bias. The nominal Hf insertion layer thickness $t_{Hf} = 0.7$ nm is chosen to reduce the damping of the free layer while maintaining the $\xi_{SH}$, as reported by Nguyen *et al*[1]. We find from flip-chip ferromagnetic resonance measurements on the exchange-biased films, however, a Gilbert damping parameter $\alpha = 0.018$ and effective magnetization $M_{eff} = 3.29 \times 10^5$ A/m which are quantitatively different from values in Nguyen *et al.*, possibly due to differences in deposition and annealing conditions. The value for $M_s t_{FeCoB}^{eff}$, the product of saturation magnetization and effective thickness of the free layer, is measured to be 0.002 A from vibrating sample magnetometry. From this, we calculate spin-Hall efficiencies in the range $\xi_{SH}^{eff} = 0.052 - 0.073$ for the three devices in this report, using the macrospin-derived relation[2] $\xi_{SH}^{eff} = \frac{2e}{\hbar} \mu_0 M_s t_{FeCoB}^{eff} \alpha \left( H_C + \frac{M_{eff}}{2} \right) / J_{c0}$. The multilayer stacks are patterned by deep-UV photolithography (ASML 300C) and etched by Ar$^+$ ion milling (IntlVac) into 335 nm wide, 600 nm long channels. Using an aligned electron beam lithography (JEOL JBX-6300FS, 100 kV) exposure and ion milling, we then define the MTJs by fabricating elliptical pillars with three different aspect ratios, as detailed in the main text, in the center of the channels. The ion-milling process is terminated when traces of the channel material become visible in the chamber's secondary ion mass spectrometry endpoint detector. After protecting the devices with electron-beam evaporated SiO$_2$, electrical connections are established to the channel and top contact of the MTJ by means of a liftoff process. The devices are annealed at 360 C for 45 minutes in a vacuum of <10$^{-6}$ Torr, in the presence of a 1.5 kG external field along their long axes.

## 2. Measurement technique.

For pulse switching experiments, two Picosecond Pulse Labs 10,070A pulse generators are routed through a voltage divider, the capacitive port of a bias-tee, and finally through microwave probes to the device leads. One of the pulse generators is used to apply rectangular switching pulses of varying amplitude $V$ and duration $\tau$ (with 65 ps rise time and 100 ps fall time), while the other is used to apply shallow reset pulses at the 10 ns maximum pulse duration. As in the *dc* ramp-rate measurements, $H_{ext}$ is adjusted to bias the samples at the centers of their magnetic-field hysteresis loops. We emphasize that this field only cancels the average projection of the reference layer's dipole field along the x direction: according to our micromagnetic simulations, some curvature of the free layer magnetization remains in the P state though it is most exaggerated in the LA devices. The initial and final MTJ resistance states are measured using a lock-in amplifier connected across a voltage divider formed by the MTJ and a 10 MΩ series reference resistor.

## 3. Micromagnetic curvature in the free layer.

Through simulations, we find the existance of micromagnetic curvature in both the AP and P states, due to the influence of the reference layer's residual dipole field, as well as due to edge roughness. Although the AP state appears to have lower curvature than the P state - due to the fact that the dipole field from the reference layer reinforces the shape anisotropy in that case - local curvature can still be expected to exist in both the AP and P states (Fig. 4a,d). Since the anisotropy field is the lowest in the LA device, we can also expect more curvature in the LA devices. This was experimentally observed from the magnetoresistance measurement comparing the three devices (Fig. 1b), where both the AP and P state resistances in the LA device vary with applied magnetic field before and after the switching points. In contrast, the AP states in the MA and HA devices show more stable resistances (Fig. 1b) beyond their respective switching fields.

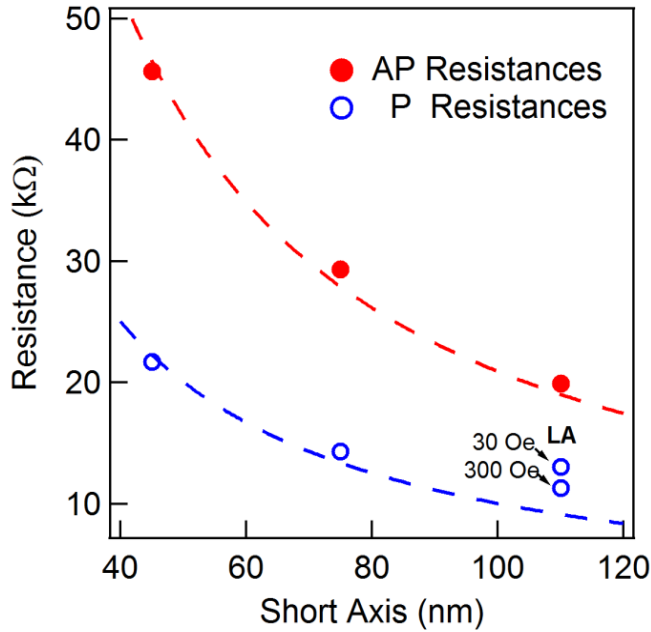

**Figure S1.** Resistance as a function of MTJ aspect ratio. The P state resistance for the low aspect (LA) ratio device shows deviation from the expected trend based on the device short axis, indicating the existence of significant curvature in the magnetization state. This deviation is reduced when the device resistance is measured at higher fields. Note that the long axis is kept constant at 190 nm for all three aspect ratios, so the smaller the aspect ratio the larger the area and smaller the resistance of the MTJ. Dashed lines are fits extrapolated from the HA and MA device resistances.

Figure S1 quantifies the raw AP and P state resistances for all three devices, with a 525 $\Omega$ contribution from the channel subtracted from the measured total MTJ resistance. Note that the spin-Hall channel contributes half of its total resistance to the measured MTJ resistance, as it forms the 'bottom lead' for the measurement circuit in the three-terminal configuration. The 'top lead' resistance is negligible as it consists of ~50 nm Ti/Pt material. Using the MA and HA device AP and P state resistances as more consistent values for comparison, it is seen that the LA device's P state resistance is 13.1 k$\Omega$ at ($H_{ext}$-$H_{off}$) of 30 Oe. (~43% higher than the expected 9.1 k$\Omega$). The magnetic origin of this behavior is is apparent in R(H) loops taken with higher fields, when the P state resistance goes down to 11.3 k$\Omega$ at ($H_{ext}$-$H_{off}$) of 300 Oe (Fig. S1). The weak anisotropy in the LA free layer likely increases the curvature in the P state.

## 4. Magnetic characterization of the free layer.

We use vibrating sample magnetometry (Quantum Design Inc) to measure the magnetic moment of an unpatterned 5 × 5 mm die from the same wafer used for fabrication of the devices. Figure S2a shows the moment per area of the free layer as a function of the externally applied magnetic field. The saturated value of $M_s \times t_{eff}$ is 0.0020 A. Using the as-deposited 1.6 nm thickness of FeCoB, we can calculate the $M_s$ to be $1.25 \times 10^6$ A/m. This value does not account for any magnet dead layer at the Hf/FeCoB interface.

We use a flip-chip technique to measure the ferromagnetic resonance in an annealed, unpatterned die from the same wafer used to fabricate the devices. Briefly, a microwave waveguide optimized for transmission in the 1-20 GHz range carries a 15 dBm rf power generated by a signal generator (Agilent E8257). The sample is placed on top of this waveguide such that the magnetic layers face the waveguide. A dc magnetic field is scanned using an external electromagnet to detect the resonance condition. This dc field is further modified by using a small ac field generated by Helmholtz coils, which provides an ac signal for lock-in detection. When the resonance condition is satisfied, microwave power is absorbed into the uniform precession mode. The changes in the absorbed power (d$P$/d$H$) are detected using a rectifying diode, at the ac field modulation frequency.

Figure S2b shows the raw data of d$P$/d$H$ versus the scanned dc magnetic field. The Lorentzian derivative fit to this data at 9 GHz frequency is also overlaid, and shows very good fidelity to the data. Figure S2c presents the variation of the fitted resonance field, as well as the Kittel model fit to this data from which we primarily obtain the value for $M_{eff}$. Finally, Fig. S2d shows the resonance linewidths as a function of the rf frequency, and the linear fit to these points gives us the magnetic damping of the free layer.

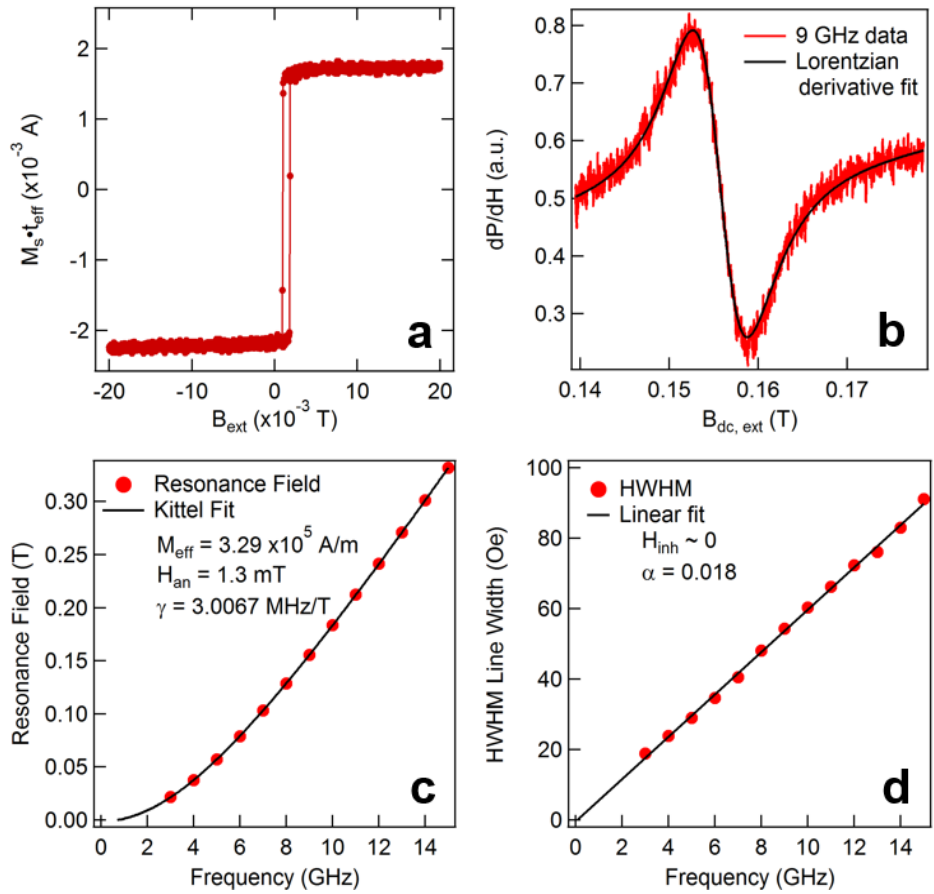

**Figure S2.** Magnetic characterization of free layer. (a) Magnetic response to an in-plane field measured using vibrating sample magnetometry. (b) A representative ferromagnetic resonance signal from the free layer measured at 9 GHz; a Lorentzian derivative fit is overlaid. (c) Fitted resonance fields as a function of applied rf frequency; a Kittel equation fit is overlaid. (d) Half width at half maximum (HWHM) line widths as a function of the rf field; a linear fit is overlaid.

## 5. Switching probability plots in the pulsed regime.

Pulsed voltage measurements are performed as described in Methods in the main text. The switching probability is measured for AP-to-P and P-to-AP polarities as a function of pulse voltage and pulse duration. For each sweep, the pulse voltage (applied with the appropriate sign for the AP-to-P or P-to-AP polarity) is set using the pulse generator's power attenuator which allows for 1 dB steps. At each combination of voltage and duration, 200 attempts are made to switch the device and the probability for switching over these attempts is recorded. Figure S3 presents the acquired data for each of the three Pt devices studied in this work. The 50% probability switching times (Fig. 2b-d) are then calculated by linear interpolation between the two nearest measured pulse durations that span the 50% probability crossing, for each pulse voltage used.

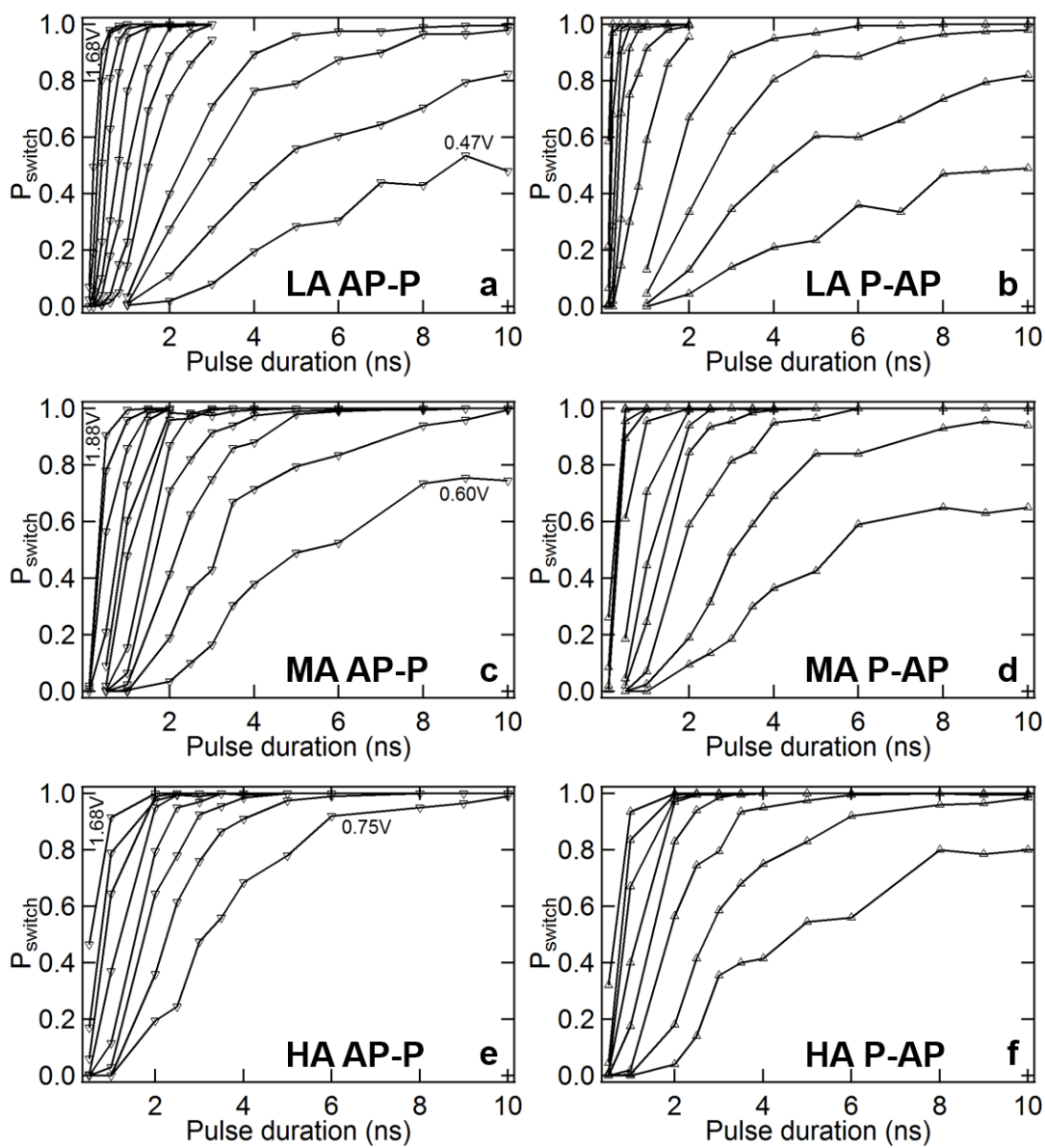

**Figure S3.** (a, c, e) AP-to-P switching probabilities for LA, MA and HA devices. (b, d, f) P-to-AP switching probabilities for LA, MA and HA devices. The minimum and maximum pulse voltages used are indicated in the plots, with intermediate voltages incrementing in steps of 1 dB in power.

## 6. Comparing Ta and Pt spin-Hall channels.

We have also carried out fast pulse switching measurements of a Ta HA three-terminal device. The tantalum materials stack is: SiOx | Ta(7) | FCB (1.8) | MgO(1.6) | FCB (3.5) | Ta (4) | Ru (4), with the numbers in parentheses indicating nanometers. The fabrication procedure is identical to the Pt device fabrication detailed in Methods. The HA device dimension is the same as the Pt HA device. The switching probability pulse sweep data is presented in Fig. S4a,b for AP-to-P and P-to-AP switching. Similar to the Pt devices, the 50% switching time can be fitted well to the macrospin model, as shown in Fig. 4c.

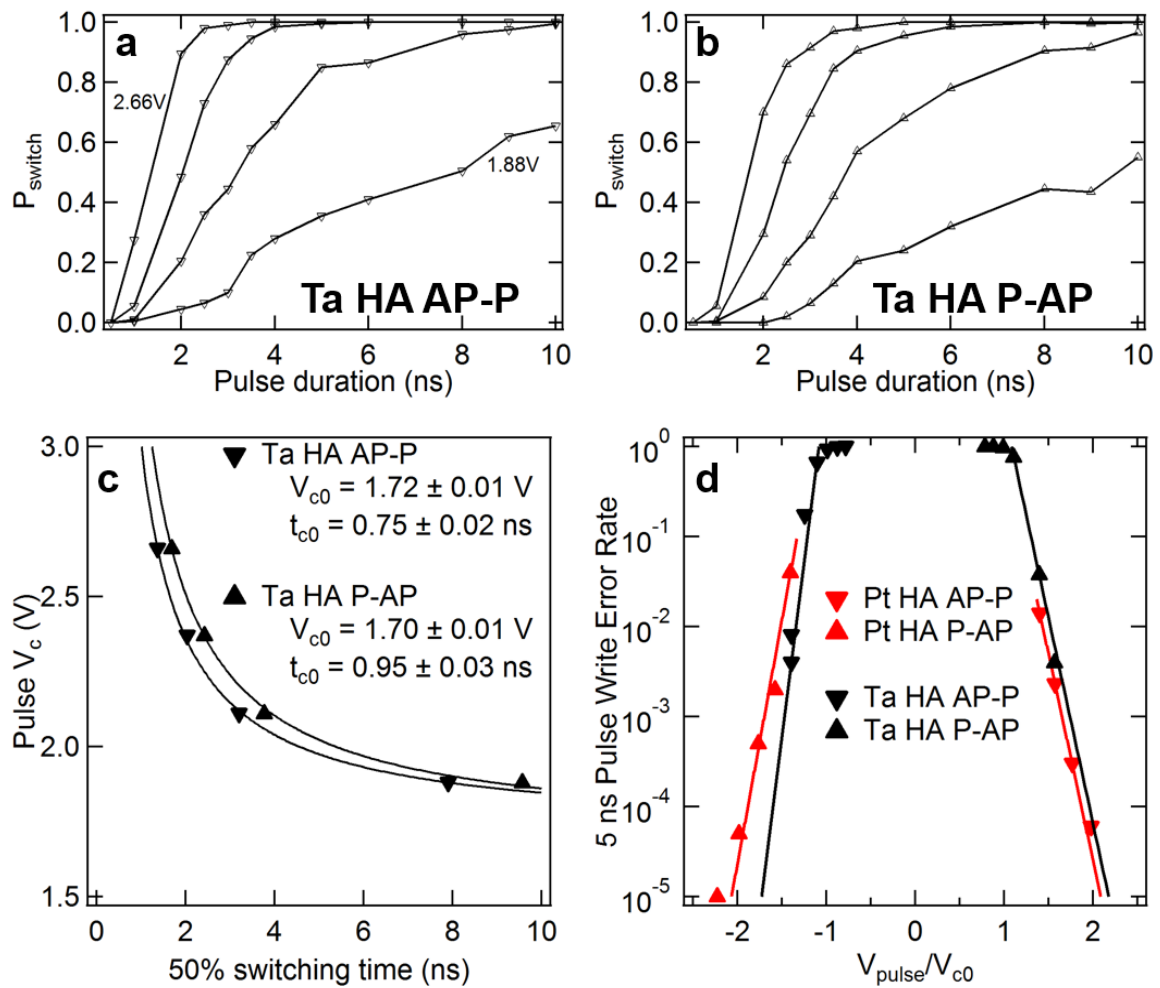

**Figure S4.** (a) AP-to-P and (b) P-to-AP pulse voltage and pulse duration sweeps. (c) Summary of switching voltage versus 50% switching times, with macrospin fits overlaid. (d) Comparison of the Ta and Pt HA device 5 ns write error rates as a function of the normalized pulsed voltages.

We note that due to the different sign of the spin Hall ratio of Ta compared to that of Pt the Oersted field points in the opposite direction for the Ta channel case compared to the Pt channel devices during switching, reinforcing the shape anisotropy. However, we observe that the $t_{c0}$ parameter is still very small (<1 ns), and comparable to the Pt devices. We do observe a slight asymmetry in the $t_{c0}$ values, with the AP-to-P switching being slightly faster than the P-to-AP switching for the Ta case. This is opposite to the Pt case where the P-to-AP switching was slightly faster for the MA and LA devices. Together, these observations indicate that the Oersted field has a more subtle role than just augmenting or diminishing the anisotropy field in the device; the highly favorable switching times appear to be a more general feature of the three terminal device geometry.

## 7. Micromagnetic simulations.

Simulations are performed using the OOMMF micromagnetic simulation package.[3] We model the switching response of MA samples, which are given a realistic edge profile (up to the in-plane spatial discretization length of 2.5 nm) taken from SEM images of our devices (Fig. S5). We use an exchange stiffness of $2\times10^{-11}$ J/m in the simulations as this value, together with the geometry, reproduces the qualitative shape of the hysteresis curve measured in the experiment. This value is close to that of Fe,[4] which is in line with the iron-rich composition of the FeCoB free layer.[5] We include both the free and reference layers in the simulation, therefore incorporating their dipolar interactions, and round their thicknesses to the nearest multiple of the 1.5 nm vertical discretization length. From the equilibrium magnetic state we apply current pulses with 65 ps rise time, including the Oersted field as a uniform magnetic field along the long axis of the sample. All simulations are performed at 0 K. Movies depicting the simulated switching mechanism for both AP-P and P-AP polarities, with and without the Oersted field, for a current density of $2.1\times10^{12}$ A/m$^2$ are available online as part of the Supporting Information.

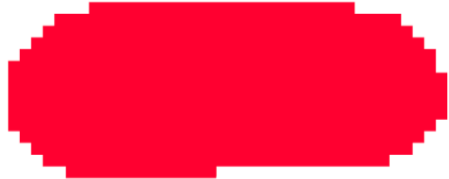

**Figure S5.** Geometry of simulated free layer, nominally an ellipse of dimension 190 × 75 nm, corresponding to the MA device in the experiments.

## 8. References.